\documentclass[sigconf]{acmart}
\AtBeginDocument{%
  }

\setcopyright{acmlicensed}
\copyrightyear{2026}
\acmYear{2026}
\setcopyright{cc}
\setcctype{by}
\acmConference[CHI EA '26]{Extended Abstracts of the 2026 CHI Conference on Human Factors in Computing Systems}{April 13--17, 2026}{Barcelona, Spain}
\acmBooktitle{Extended Abstracts of the 2026 CHI Conference on Human Factors in Computing Systems (CHI EA '26), April 13--17, 2026, Barcelona, Spain}
\acmDOI{10.1145/3772363.3799082}
\acmISBN{979-8-4007-2281-3/2026/04}

\usepackage{subcaption}
\usepackage{listings}
\begin{document}

\title[\textsc{Synonymix}]{\textsc{Synonymix}: Unified Group Personas for Generative Simulations}

\author{Huanxing Chen}
\authornote{Both authors contributed equally to this research.}
\email{huanxing@stanford.edu}
\orcid{0009-0005-0895-3038}
\affiliation{%
  \institution{Stanford University}
  \city{Stanford}
  \state{CA}
  \country{USA}
}

\author{Aditesh Kumar}
\authornotemark[1]
\email{aditesh@stanford.edu}
\orcid{0009-0003-6466-1211}
\affiliation{%
  \institution{Stanford University}
  \city{Stanford}
  \state{CA}
  \country{USA}
}

\renewcommand{\shortauthors}{Chen and Kumar}

\begin{abstract}

  Generative agent simulations operate at two scales: individual personas for character interaction, and population models for collective behavior analysis and intervention testing. We propose a third scale: {meso-level simulation} - interaction with group-level representations that retain grounding in rich individual experience. To enable this, we present \textsc{Synonymix}, a pipeline that constructs a "unigraph" from multiple life story personas via graph-based abstraction and merging, producing a queryable collective representation that can be explored for sensemaking or sampled for synthetic persona generation. Evaluating synthetic agents on General Social Survey items, we demonstrate behavioral signal preservation beyond demographic baselines (p<0.001, r=0.59) with demonstrable privacy guarantee (max source contribution <13\%). We invite discussion on interaction modalities enabled by meso-level simulations, and whether "high-fidelity" personas can ever capture the texture of lived experience.
\end{abstract}



\begin{teaserfigure}
  \includegraphics[width=\textwidth]{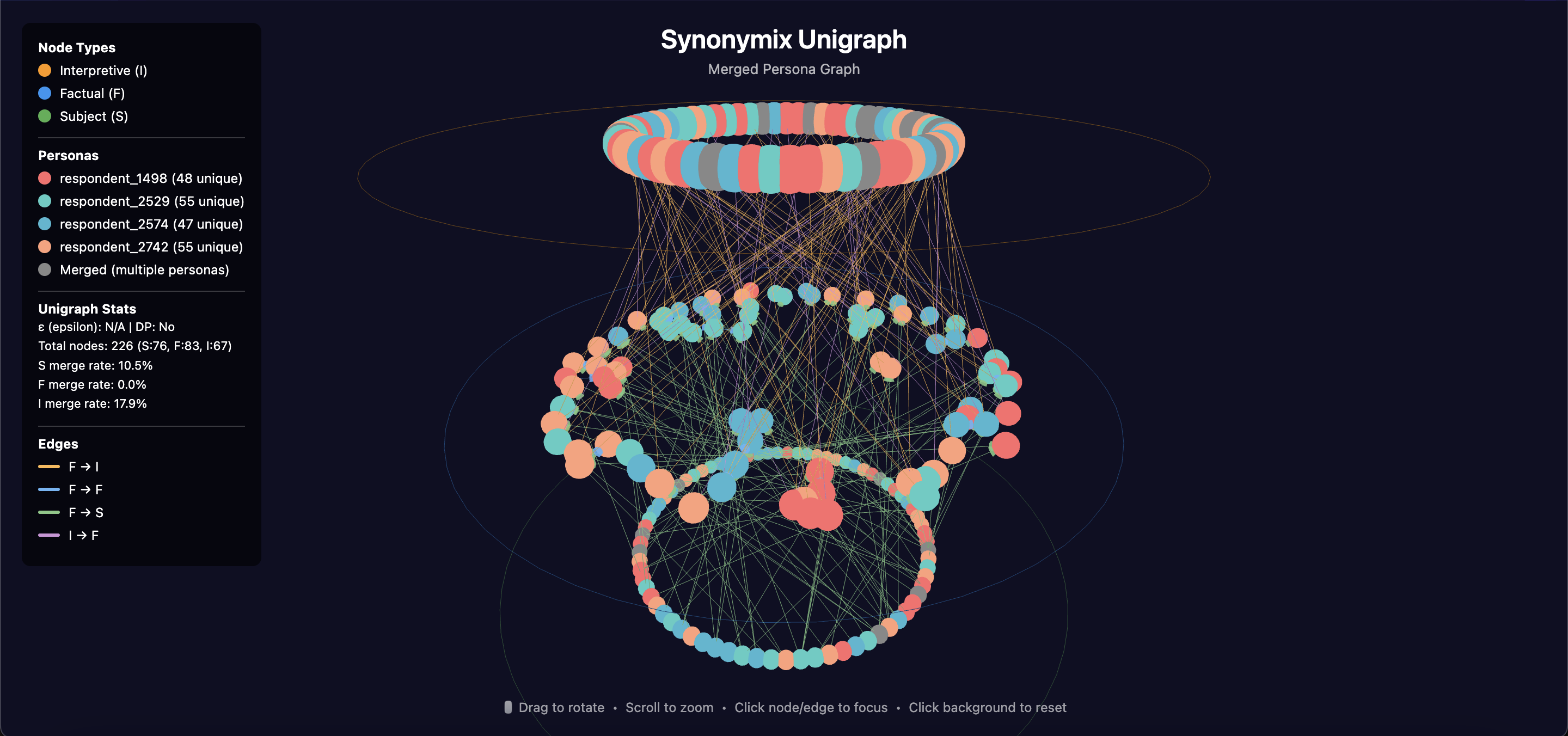}
  \caption{The central artifact of \textsc{Synonymix} is the "unigraph", a semantically unified knowledge graph constructed from each participant's narrative persona. This artifact, which captures the shared personal history of the collective, can be explored for greater understanding or sampled for synthetic representation, and offers a discrete structure upon which formal privacy guarantees can be introduced for the participants. It is visualized here with $n=4$ for readability.}
  \Description{This figure depicts the central artifact of \textsc{Synonymix}, which is a 3-tiered graph. The three layers, from bottom to top, are comprised of subject nodes, factual nodes, and interpretive nodes respectively. Each node is colored according to the participant from which it originated, and linked to nodes from all similar concepts across the entire agent bank. There is a sidebar for this simple visualization with $n=4$, which notes categories for the colors: respondent_1498 (48 unique), respondent_2529 (55 unique), respondent_2574 (47 unique), respondent_2742 (55 unique), Merged (multiple personas). It also unigraph stats: \epsilon (epsilon): N/A | DP: No; Total nodes: 226 (S:76, F:83, I:67); S merge rate: 10.5\%; F merge rate: 0.0\% I merge rate: 17.9\%}
  \label{fig:teaser}
\end{teaserfigure}


\maketitle
\section{Introduction \& Background}

Plausible models of human behavior are an elusive and yet coveted construction throughout the history of the cognitive sciences. They are particularly compelling for their utility in simulation, displaying a unique ability to surface insights into individual and collective behavior. Historical exemplars include Card et al.'s Model Human Processor for individual-level interaction \cite{card1986model} and Schelling's agent-based models demonstrating emergent collective segregation patterns \cite{schelling1971dynamic}. Both works inspired a body of literature fascinated with understanding human behavior through simulation.

This history of work is now joined by generative agent-based models (GABMs) \cite{ghaffarzadegan2024generative, larooij2025large, park2023generative}, which leverage language models to construct believable proxies of human behavior. However, if we conceive simulation's essence as the \textit{exploration} of counterfactual outcomes and selves, then the representational adequacy of GABMs warrants scrutiny. LLM agents struggle with faithful representation of different perspectives from demographic variables alone \cite{cheng2023compost, batzner2025whose, naous2023having} and do not easily capture the contextual particularity of individual experience. Recognizing that thin persona specifications yield agents that poorly approximate the heterogeneity of real social actors, studies have begun exploring more high-fidelity simulations by seeding generative agents with personas derived from life story interviews \cite{park2024generative}. With life stories being a particularly rich characterization, a core tension arises: the same biographical depth that enables higher-fidelity simulation also heightens privacy risks for participants.

We ask: \textit{can we find a high-fidelity yet privacy-preserving representation of an agent bank?} Synthetic personas are a prominent approach, with recent work exploring fine-grained synthesis at the population \cite{castricato-etal-2025-persona} and individual \cite{jandaghi-etal-2024-faithful} levels. However, both approaches extrapolate from a seed dataset of sampled features and face additional challenges in representing the rich narratives of their source participants. Rather than synthesizing individuals or populations, we reframe the challenge as producing synthetic agent banks that faithfully and privately represent the fine-grained narratives of personas in the agent bank as a whole. We propose abstracting one level higher from the individual to the \textit{group} as a means of easing the fidelity-privacy tension - a collective representation grounded in individual-level nuance, but abstracted enough to be safely shareable. This reframing surfaces a second question: if such a "high-fidelity" group-level artifact were possible, what interaction modalities might it enable beyond individual-level character interaction or population-level aggregate behavior? We term this space \textit{meso-level simulation}.

We raise \textsc{Synonymix} as a potential answer to these questions. \textsc{Synonymix} unifies high-fidelity personas by extracting knowledge graphs from each and merging them according to \textit{synonymity} to produce a unified graph ("unigraph"). The unigraph can be constructed in a privacy-preserving way and sampled for synthetic persona generation. More importantly, we suggest that this unigraph constitutes a novel artifact of collective representation that could be explored interactively for sensemaking and other meso-level interactions that retain the essence of simulation - \textit{exploration}.

\section{Methodology}

\subsection{Life Story Graph Extraction}
We define a persona graph ontology for decomposing textual life stories into structured representations. The schema specifies allowable node types, edge types, and edge labels.

\textbf{Node Types:} The ontology includes three node types: \textit{Subject nodes (S)} represent recurring proper nouns (people, places, institutions); \textit{Factual nodes (F)} capture concrete events, actions, or milestones;  and \textit{Interpretive nodes (I)} encode values, motivations, or reflective self-narratives derived from factual experiences. Inspired by Park et al.'s use of expert reflection module involving simulated experts from 4 branches of social sciences \cite{park2024generative}, I-nodes are generated via LLM annotation simulating expert perspectives from psychologists, sociologists, and anthropologists.

\textbf{Edge Grammar:} Edges are restricted to four types: \textit{F→S} (spatial, temporal, and relational markers), \textit{F→F} (temporal or causal relations between events),  \textit{F→I} (derived meanings from events), and  \textit{I→F} (values influencing actions). For F→S edges, we role-type participants using an inventory adapted from PropBank \cite{palmer2005proposition} and FrameNet \cite{baker1998berkeley} for the autobiographical domain (see Appendix \ref{edge_grammar} for full edge inventory).

Certain edge types are deliberately excluded: S→F (redundant with role-typed F→S), I→S (values are always mediated through factual experiences), and I→I (prevents unbounded chaining of abstract meanings). This maintains balance between expressive power and structural tractability.

\subsection{Privacy-Preserving Graph Aggregation}
Individual persona graphs are merged into the unigraph through two operations. First, \textit{label genericization}: factual nodes are converted to their generic variant with non-generic entities extracted and connected via an F->S edge (e.g., F: "Graduated from Harvard University" → F: "Graduated from University" + S: "Harvard University") to reduce direct identifiability while preserving semantic relationships. Second, \textit{node merging}: nodes with identical or semantically equivalent labels across personas are merged, with provenance tags tracking source contributions. This creates shared "hubs" (e.g., a generic "Father" S-node, or generic "Economic Insecurity Driving Academic Achievement" I-node connecting multiple personas' experiences) that enable cross-persona traversal while obscuring individual trajectories. Notably, the discrete graph structure supports \textit{differentially private (DP)} \cite{dwork2006calibrating, dwork2014dpfoundations} merging\footnote{We construct the unigraph as a private histogram of nodes using a differentially private set union \cite{gopi2020dpsu1, gopi2020dpsu2}, which prunes nodes that are underrepresented in agent bank to avoid leakage.}, though we defer systematic DP evaluation to future work given the node sparsity challenges at our limited agent bank size of N=30.

\begin{figure*}[h]
    \centering
    % --- 1st Subfigure ---
    \begin{subfigure}[t]{0.48\textwidth}
        \setlength{\fboxsep}{0pt}
        \colorbox[HTML]{0F0F22}{
        \parbox[b][5cm][t]{\linewidth}{
            \centering
            \includegraphics[width=\linewidth, height=5cm, keepaspectratio]{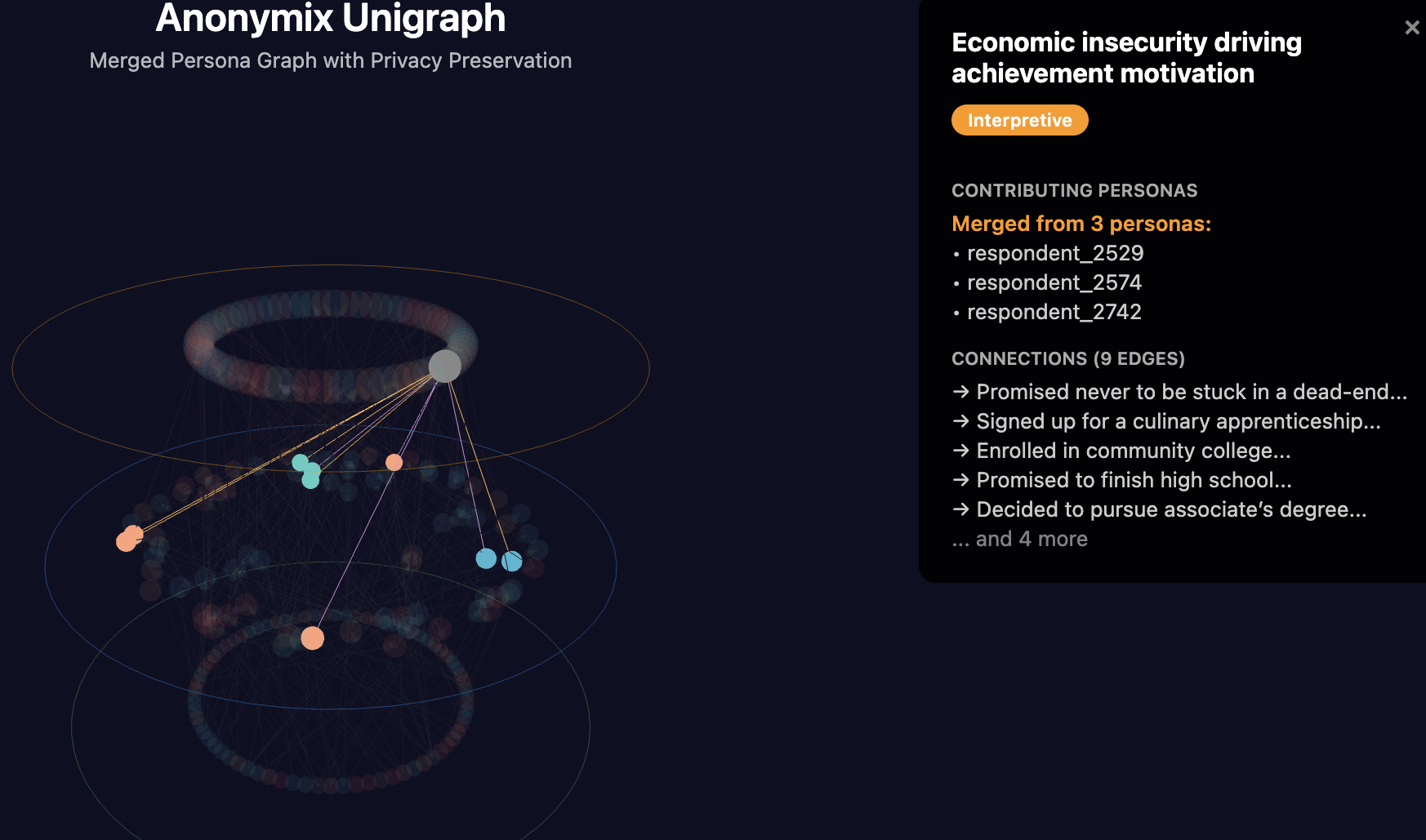}}}
        \Description[Unigraph with I-node selected: 'Economic insecurity driving academic achievement']{Unigraph visualization which highlights all adjacent nodes for a selected node. An Interpretive node 'Economic insecurity driving academic achievement' is selected from the top layer (I) of the graph, with the visualization noting the node was merged from 3 personas (respondents 2529, 2574, and 2742). There are 9 connections to nodes about: 'Promised to never be stuck in a dead-end...', 'Signed up for a culinary apprenticeship...', 'Enrolled in community college..', 'Promised to finish high school...', Decided to pursue associate's degree...', and 4 more not displayed.}
        \caption{I-node: Economic insecurity driving academic achievement.}
        \label{fig:grid1}
    \end{subfigure}
    \hfill 
    % --- 2nd Subfigure ---
    \begin{subfigure}[t]{0.48\textwidth}
        \setlength{\fboxsep}{0pt}
        \colorbox[HTML]{0F0F22}{
        \parbox[b][5cm][t]{\linewidth}{%
            \centering
            \includegraphics[width=\linewidth, height=5cm, keepaspectratio]{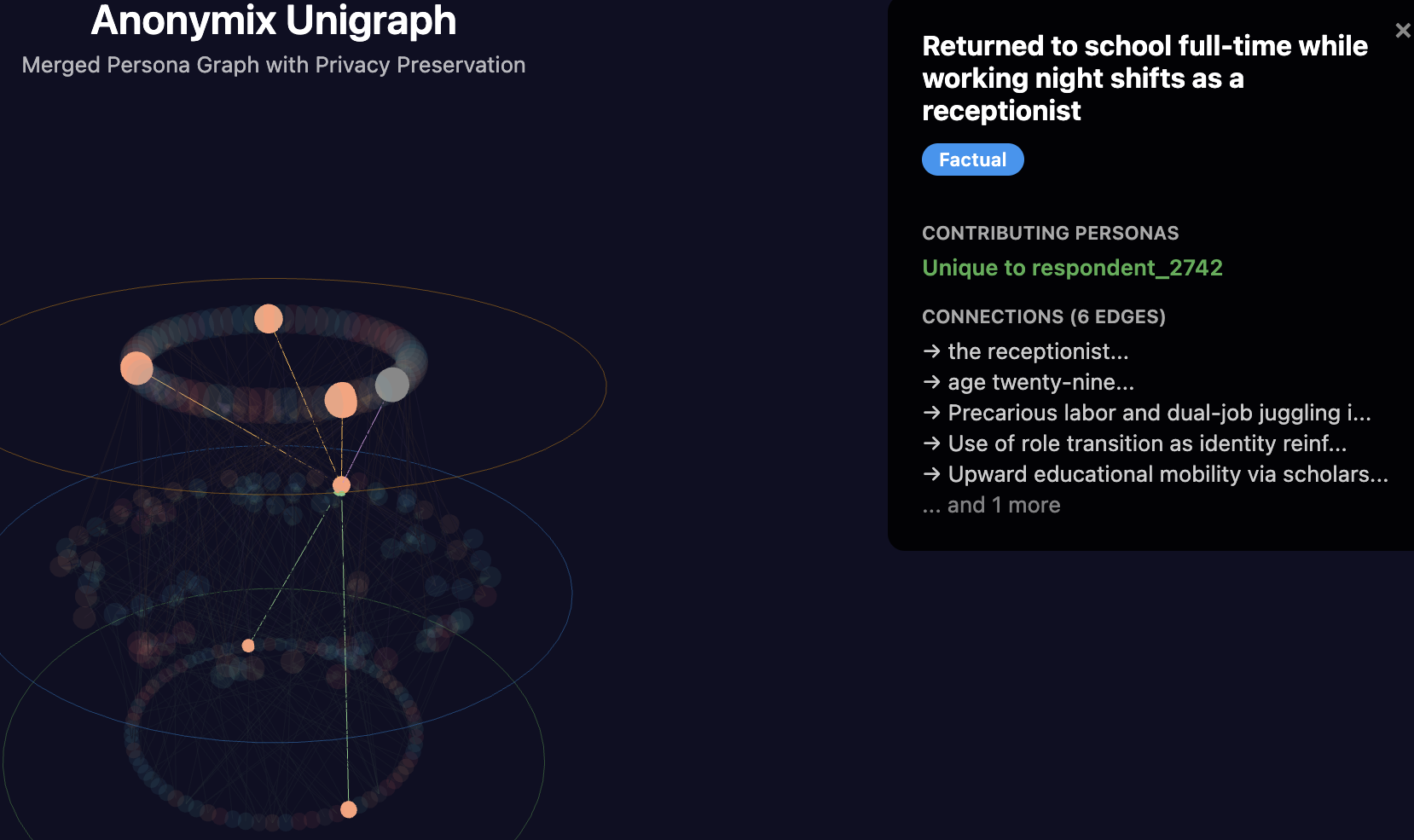}}}
        \Description[Unigraph with F-node selected from Respondent 2742.]{Same unigraph visualization as Fig 2a, where the adjacent Factual node 'Returned to school full-time while working night shifts as a receptionist' is selected from the middle layer (F) of the graph. The node is unique to respondent 2742 and has 6 connections to nodes about: 'the receptionist...', 'age twenty-nine...', 'Precarious labor and dual-job juggling i...', 'Use of role transition as identity reinf...', 'Upward educational mobility via scholars...', and 1 more not displayed.}
        \caption{Persona 1: I-node reflected in night shifts during schooling.}
        \label{fig:grid2}
    \end{subfigure}
    \\[\baselineskip]

    % --- 3rd Subfigure ---
    \begin{subfigure}[t]{0.48\textwidth}
        \setlength{\fboxsep}{0pt} 
        \colorbox[HTML]{0F0F22}{
        \parbox[b][5cm][t]{\linewidth}{
            \centering
            \includegraphics[width=\linewidth, height=5cm, keepaspectratio]{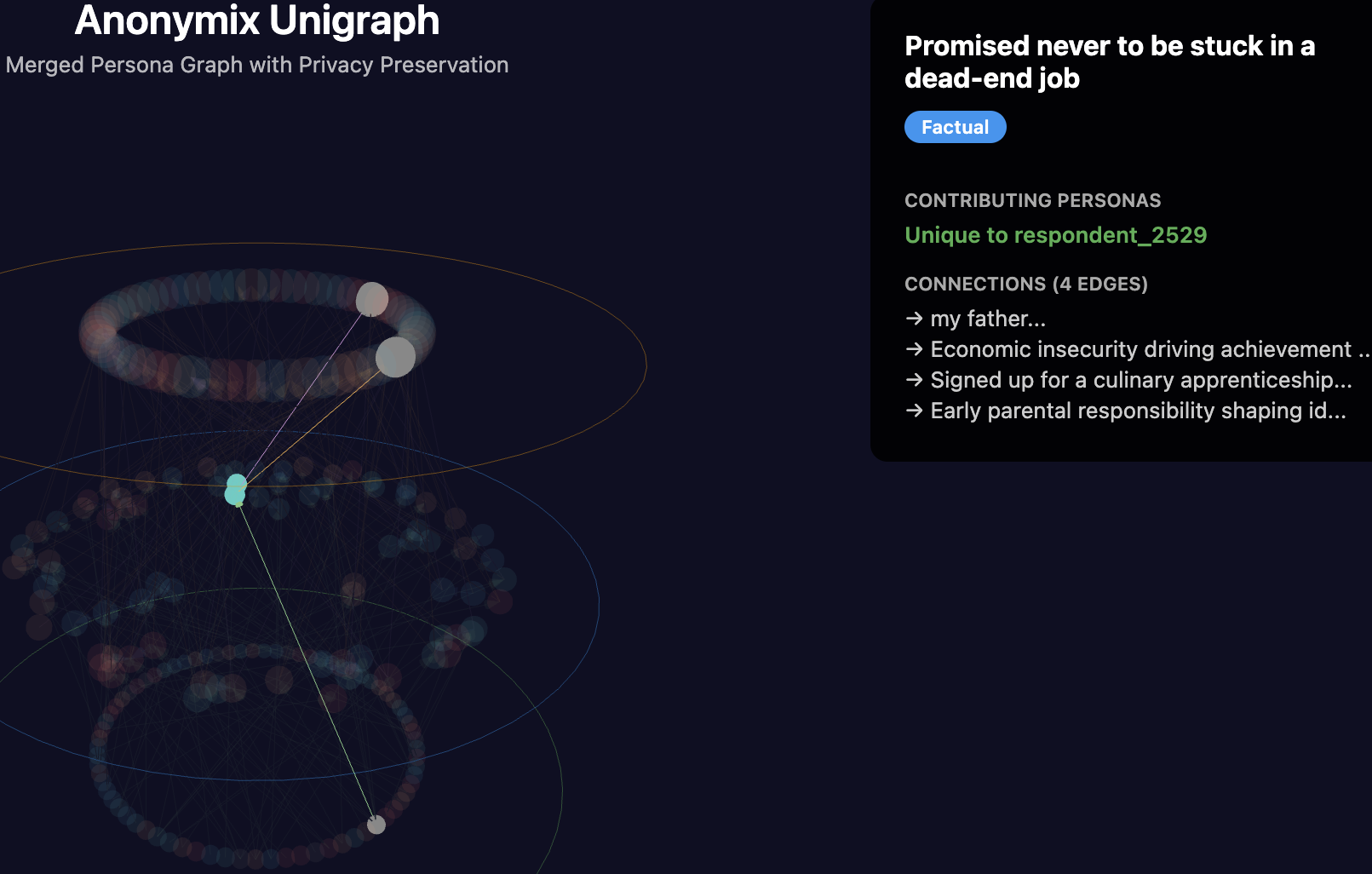}}}
        \Description[Unigraph with F-node selected from Respondent 2529.]{Same unigraph visualization as Fig 2a, where the adjacent Factual node 'Promised to never be stuck in a dead-end job' is selected from the middle layer (F) of the graph. The node is unique to respondent 2529 and has 4 connections to nodes about: 'my father...', 'age Economic insecurity...', 'Signed up for a culinary apprenticeship...', and 'Early parental responsibility shaping id...'}
        \caption{Persona 2: Same I-node reflected in promise to father to never be stuck in a dead-end job.}
        \label{fig:grid3}
    \end{subfigure}
    \hfill
    % --- 4th Subfigure ---
    \begin{subfigure}[t]{0.48\textwidth}
        \setlength{\fboxsep}{0pt}
        \colorbox[HTML]{0F0F22}{
        \parbox[b][5cm][t]{\linewidth}{
            \centering
            \includegraphics[width=\linewidth, height=5cm, keepaspectratio]{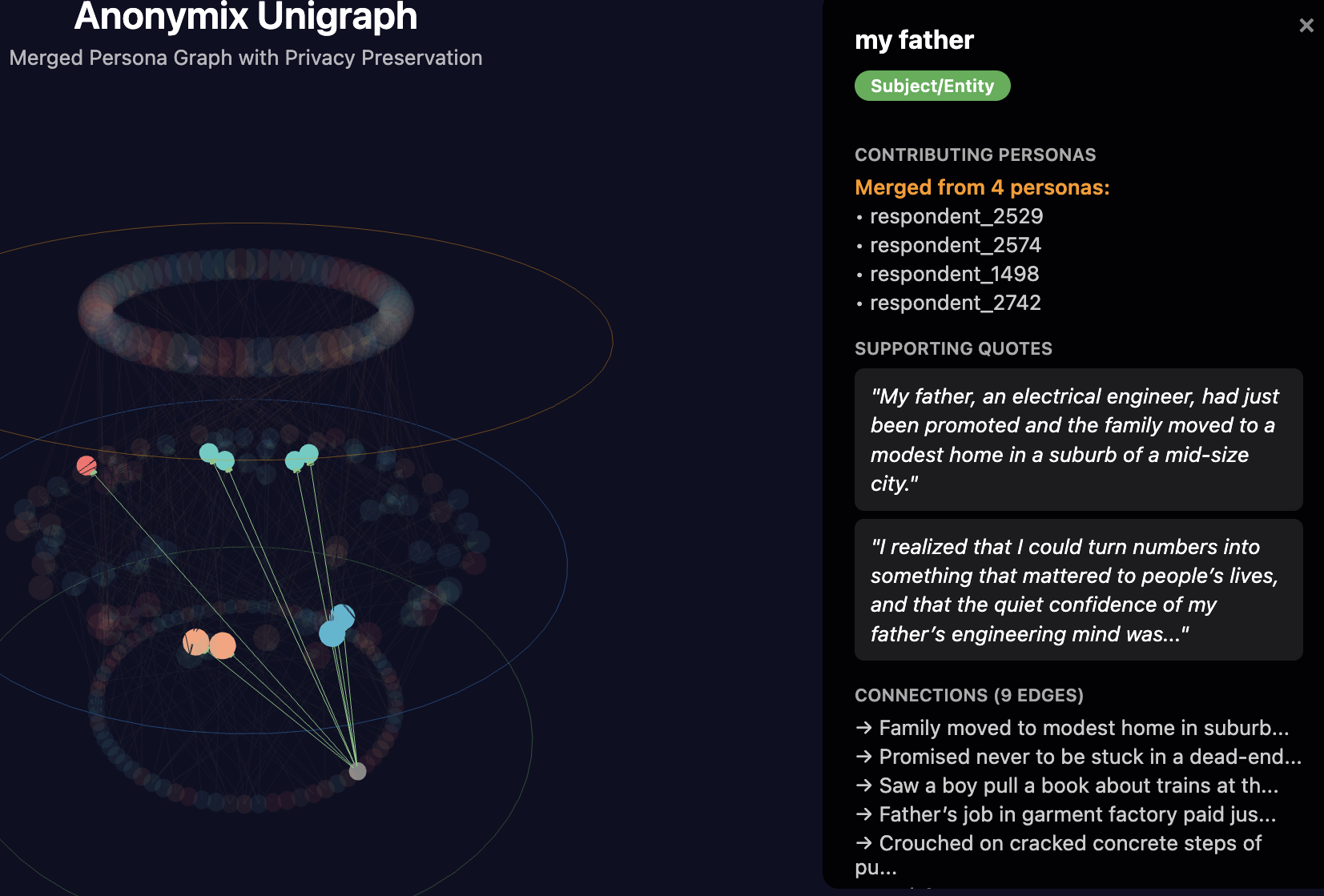}}}
        \Description[Unigraph with S-node selected: 'my father'.]{Same unigraph visualization as Fig 2a, where the adjacent Subject/Entity node 'my father' is selected from the bottom layer (S) of the graph. The node is merged from the 4 personas (respondent 2529, 2574, 1498, and 2742). The node is noted to have two supporting quotes: "My father, an electrical engineer, had just been promoted and the family moved to a modest home in a suburb of a mid-size city" and "I realized that I could turn matters into something that mattered to people's lives, and that quiet confident of my father's engineering mind was...". The node has 9 connections to nodes about: 'Family moved to a modest home in a suburb...', 'Promised never to be stuck in a dead-end...', 'aw a boy pull a book about trains at th...', 'Father's job in garment factory paid jus...', and 'Crouched on cracked concrete steps of pu...'.}
        \caption{S-node hub (Father) connecting personas and events.}
        \label{fig:grid4}
    \end{subfigure}

    \caption{Unigraph Traversal on 4 Personas}
    \label{fig:unigraph_traversal}
\end{figure*}

\subsection{Graph Traversal Sampling}

After constructing a unigraph of merged persona graphs, we adopt a variant of the Random Walk with teleportation traversal strategy \cite{leskovec2006sampling} to sample novel persona graphs. To reduce the risk of sampling incoherent or implausible personas, we introduce \textit{Thematic Random Walk} which maintains a 'theme anchor' during traversal represented as a vector embedding of a chosen I-node. We use I-nodes as anchors because they provide abstraction over heterogeneous events: different source personas may have factual nodes that instantiate the same underlying theme (I). Thus, anchoring to interpretations allows recombination of events across sources while maintaining motivational coherence, and we instantiate the strength of thematic anchoring as a tunable parameter in graph traversal.

\section{Results and Analysis}
Our evaluation uses LLM-generated life story personas to establish internal validity: whether the \textsc{Synonymix} pipeline preserves sufficient behavioral signal in source personas. The algorithmic nature of the pipeline operations suggests that they should transfer to real-world data, though empirical validation remains future work.

\subsection{Data Preparation}
We construct three agent banks (N=30) from demographic seeds drawn from the General Social Survey 2022 \cite{GSS2022}, a longitudinal U.S. survey tracking attitudes on politics, religion, and social trust that is widely used in social scientific research.

\textbf{{Demographic Agent Bank (D)}}: Agents initialized with a concatenated string of randomly-sampled respondents' available demographic seeds from a set of 68 questions.

\textbf{{Life Story Agent Bank (L)}}: Narrative personas generated from the demographic seeds using McAdams' life story interview framework, a foundational instrument in narrative psychology for eliciting identity-shaping experience \cite{mcadams2008life}.  

\textbf{{Frankenstein Agent Bank (F)}}: Synthetic personas produced by \textsc{Synonymix} pipeline: life graphs are extracted from L bank personas, merged into a unified graph structure, "Frankenstein" graphs are generated via graph traversal algorithms, and these are used to reconstruct coherent textual narratives via a McAdams prompt.

Refer to Appendix \ref{personas} for example personas, and Appendix \ref{persona_method} for discussion on methods used to generate life story personas.

\subsection{Evaluation Design}

\subsubsection{Evaluating Behavioral Signal Preservation} 
We evaluate all three agent banks on a filtered set of 108 non-demographic items from the GSS 2022 Core questionnaire comprising both ordinal and nominal questions (See Appendix \ref{GSS_2022} for item examples). We're interested in using population-level distributional distance as a proxy for a simulation's fidelity as \textsc{Synonymix}'s F agent bank lacks a ground truth for individual-level metric evaluation. 

For each of the ordinal items ($n=69$), we calculate the \textit{Earth Mover's Distance (EMD)}, a metric measuring the minimum "work" required to transform one distribution into another while respecting ranked ordering. For nominal items ($n=39$), we calculate the \textit{Total Variation Distance (TVD)}, which measures how much the two distribution disagree overall when treating category mismatches equally. 

Using the corresponding measure for each item type, we calculate the following distance metrics per question:
\begin{itemize}
    \item \textbf{Enrichment distance, $\text{dist}(D, L)$}: The distance between the response distribution of all D bank and L bank agents. Represents the behavioral signal added by narrative expansion
    \item \textbf{Transformation distance, $\text{dist}(L, F)$}: The distance between the response distribution of all L bank and F bank agents. Represents the behavioral signal lost by the \textsc{Synonymix} pipeline.
\end{itemize}

Our primary hypothesis is that the average {transformation distance < enrichment distance} across all items, which would indicate that \textsc{Synonymix} pipeline \textit{loses less behavioral signal than that gained by narrative expansion from demographic personas}. This would validate graph-based transformation as a viable approach for creating high-fidelity agent banks that preserve meaningful behavioral grounding while mitigating re-identification risk.

We test this hypothesis using a one-sided Wilcoxon signed-rank test, a non-parametric paired test appropriate for comparing distances within items without distributional assumptions. We report rank-biserial correlation ($r$) as effect size and interpret $r>0.5$ as large and $r>0.3$ as medium.

\subsubsection{Evaluating Privacy}
To evaluate the degree of privacy guarantee provided by \textsc{Synonymix}, we examine \textit{Maximum Source Contribution (MSC)}, the empirical measure of the proportion of a synthetic persona's nodes attributable to its most dominant source individual. Using fractional attribution (merged nodes split credit equally among contributing sources), MSC quantifies whether synthetic personas represent genuine re-combinations or lightly-perturbed copies. We adopt $\text{MSC} < 0.50$ as our threshold: no single source should contribute a majority of any synthetic persona's content. 

\subsection{\textsc{Synonymix} Persona Performance}
\begin{figure}[h!]
    \centering
    % --- LEFT GRAPH ---
    \begin{subfigure}[b]{0.45\textwidth}
        \centering
        \includegraphics[width=\textwidth]{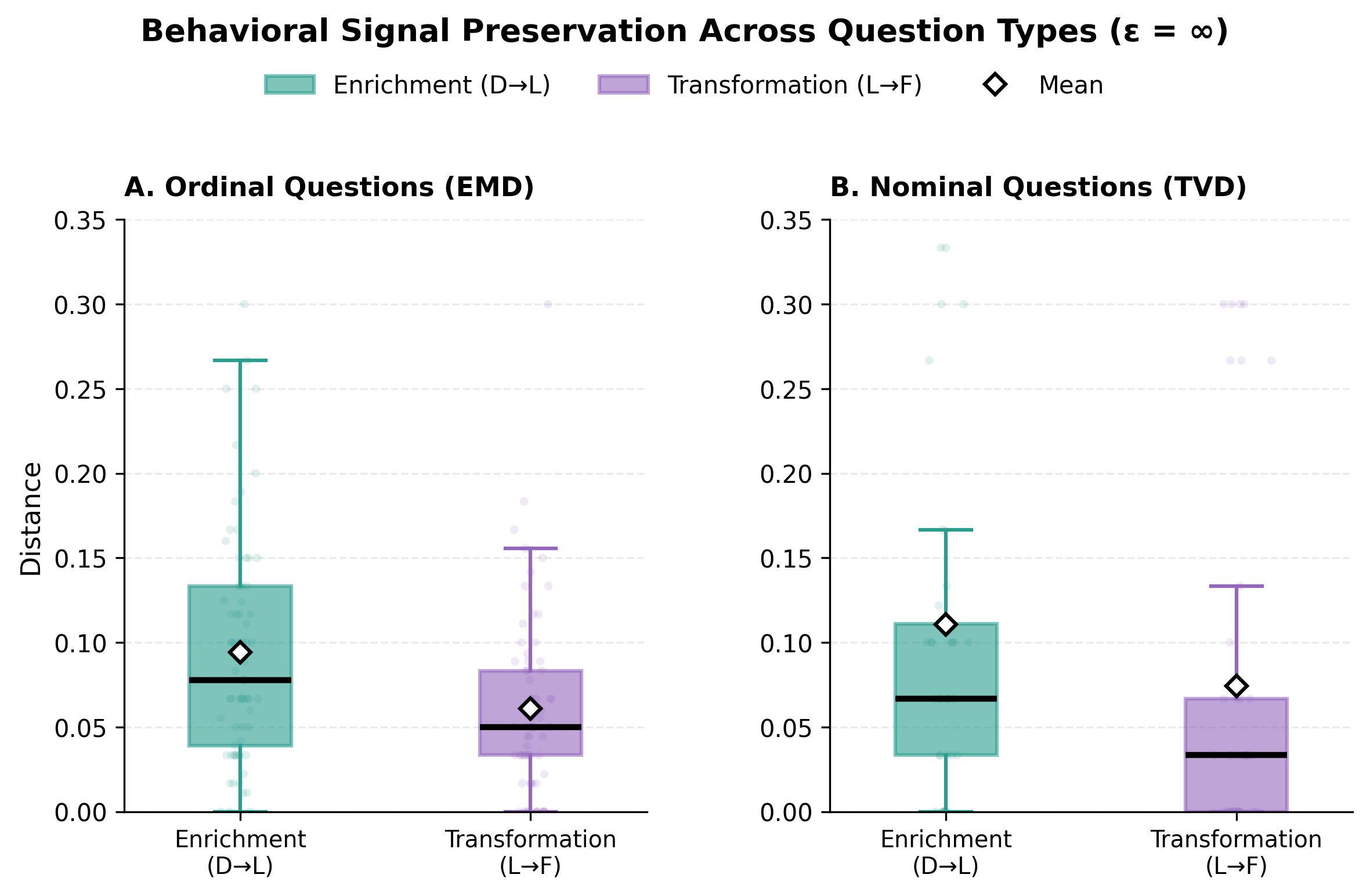}
        \Description[Enrichment and transformation distance boxplots]{Side-by-side box plots indicating the median and mean of enrichment and transformation distance, showing that enrichment distance exceeds transformation distance in the case where no privacy-preservation (differential privacy) are applied ($\epsilon = \infty$). Comparisons for ordinal questions (EMD) and nominal questions (TVD) are displayed as two separate figures.}
        \caption{Aggregate enrichment and transformation distance}
        \label{fig:behavior_left}
    \end{subfigure}
    \hfill % Pushes images apart
    % --- RIGHT GRAPH ---
    \begin{subfigure}[b]{0.45\textwidth}
        \centering
        \includegraphics[width=\textwidth]{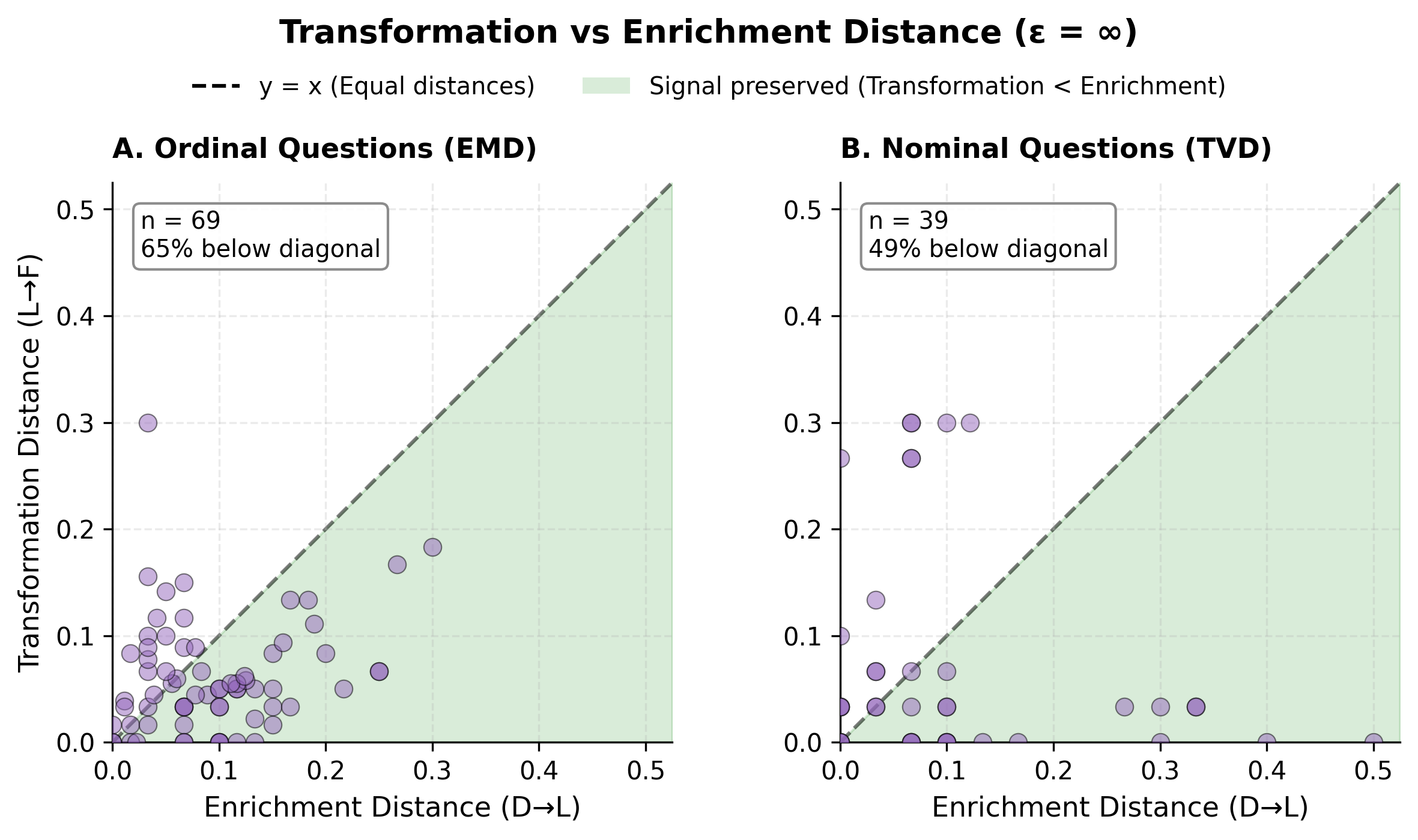}
        \Description[Enrichment and transformation distance scatterplot by question]{Scatter plots of enrichment distance (x) against transformation distance (y), where the area underneath the line y=x is shaded green to indicate enrichment exceeds transformation. For ordinal questions (END), 65\% of points lie below the line, whereas for nominal questions (TVD), 49\% of questions lie below the line.}
        \caption{Per-question enrichment and transformation distance}
        \label{fig:behavior_right}
    \end{subfigure}
    
    \caption{\textbf{TOP:} Box-plot comparison of aggregate enrichment (D→L) and transformation (L→F) distances across all GSS items. Diamonds indicate means; horizontal lines indicate medians. Two enrichment outliers ($0.4$, $0.5$) for nominal questions not shown for scale. \textbf{BOTTOM:} Per-item comparison of transformation vs. enrichment distance. Points below the diagonal indicate items where the pipeline loses less signal than narrative expansion adds.}
    \label{fig:behavioral_signal}
\end{figure}

\subsubsection{Behavior Signal Preservation}
Figure \ref{fig:behavioral_signal} presents the core comparison across all ordinal and nominal questions respectively. For ordinal questions, transformation distance was significantly lower than enrichment distance ($\text{dist}(L,F)=0.061 < \text{dist}(D,L)=0.094$), supporting our hypothesis (Wilcoxon W, $p<0.001$, $r=0.585$, large effect). $65\%$ of ordinal questions showed transformation distance below enrichment distance. For nominal questions, the pattern held directionally ($\text{dist}(D,L)=0.074 < \text{dist}(L,F)=0.111)$ with medium effect size ($r=0.462$), but did not reach statistical significance ($p=0.103$) likely due to limited statistical power ($n=39$ vs. $n=69$). Only $49\%$ of nominal questions showed transformation distance below enrichment distance.

\subsubsection{Privacy Preservation}
Across all F bank personas, MSC remained well below threshold (mean=0.129, SD=0.031, range: 0.091–0.195), with 100\% of personas below 0.50. On average, each synthetic persona drew from 29.4 of 30 source individuals. These results confirm that \textsc{Synonymix} produces genuine re-combinations rather than identifiable copies of source personas.

\section{Discussion}

\subsection{Generating Synthetic, High-Fidelity Agent Banks from Grounded Data}

\textsc{Synonymix} enables the creation of high-fidelity synthetic life story agent banks grounded in real-person data while preserving privacy through the affordances of symbolic graph structures (node merging, label genericization, potential differential privacy). While we used life story personas as proof of concept, graph-based abstraction offers a generalizable pattern for privacy-preserving synthetic data generation in domains with inherent relational structure. Practitioners can generate synthetic personas beyond the original dataset size through graph traversal, thereby addressing a bottleneck in LLM-based high-fidelity agent generation: the tendency to produce flattened demographic caricatures \cite{cheng2023compost}.

A potential caution, however,  pertains to the risk of generating incoherent or implausible life stories. While we tried to mitigate this risk through \textit{Thematic Anchoring}, the nature of probabilistic graph traversal implies the possibility of generating improbable personas who might, for instance, be a "chronically-ill 90 year-old" that "enjoys tennis". Although one could argue that such individual-level implausibilities matter less when studying aggregate, population-level dynamics, future work should explore additional coherence constraints to broaden the pipeline's applicability. In particular, ensuring the preservation of the unique lived experiences of individuals with intersectional identities remains a critical area for refinement.

\subsection{Ethical Considerations \& Future Validation}

\textsc{Synonymix} interfaces with the competing values of integrity and privacy. Due to ethics-related concerns about using high-fidelity human data in an unvalidated pipeline, we experiment only with synthetic agent banks, with findings that confirm internal validity alone. Probabilistically-sampled narratives do not capture the nuances of real human personas, raising concerns about consistency, granularity, and representation of intersectional identities. Nonetheless, our results show initial promise motivating empirical evaluation on real-human data - exploring whether the pipeline can faithfully capture persona richness while defending against structural \cite{narayanan2008robust} and LLM-driven \cite{lermen2026large} de-anonymization attacks.

\subsection{Meso-Level Interactions}
Beyond synthetic persona generation, the unigraph itself constitutes an interactive artifact - a queryable, high-fidelity representation of a collective that preserves individual nuance while remaining shareable without compromising privacy. This enables what we propose as the 'meso-level' interactions of generative agent simulations: simulations at a scale between individual-level AI personas \cite{morris2025generative} and population-level modeling \cite{chopra2025large}. At this scale, we can explore the utility of simulation beyond predictive accuracy or believability, thereby opening up a design space for simulations as a substrate for the exploration of counterfactual selves and collectives at the level of the group. 

We identify two potential applications from our prototype implementation. First,  \textit{cross-temporal and cross-demographic sensemaking}: constructing  "group personas" (e.g., "the Class of 1985") to explore how the collective experienced historical moments through navigating shared S-node hubs, as well as exploring the tension between I-node thematic similarities and the associated F-node lived experiences. And second, \textit{consensus building}: using the graph to identify convergent and divergent interpretations within stakeholder groups to support collaborative deliberation. Figure \ref{fig:unigraph_traversal} illustrates an example of the former. More broadly, treating the group as a representational unit surfaces a design space not just for within-group sensemaking, but for \textit{between-group} dynamics - exploring how collectives differ and converge in ways that are invisible at the individual scale and collapse into aggregate noise at the population scale.

What unifies these applications is their orientation rather than their scale - a mode of engagement better suited to open-ended inquiry than to prediction or output. Zhang's framing of the non-consequential and the dialectical in HCI is instructive here \cite{zhang2024searching}: interactions that are valuable not for the answers they produce but for the thinking they provoke, and that unfold through tension and exchange rather than resolution. This, we suggest, is the distinctive promise of meso-level simulation and the sensibility that should inform its design.

\subsection{Ontologies of Generative Agent Simulations}
The design choices embedded in \textsc{Synonymix} make visible something that has received little explicit attention in the generative agent literature: that every simulation encodes an ontology of personhood. What is a person? What counts as a unit of memory? Is a person defined by a stable set of attributes, or are they constituted through their relations and social roles? These are not technical questions, and different answers open up fundamentally different design spaces.

Current generative agent architectures tend to inherit the assumptions of methodological individualism: agents as discrete units with stable attribute bundles, and collectives as aggregates \textit{of}  individuals. But this is one ontological choice among many. Emirbayer's relational sociology, for instance, proposes foregrounding the relations over the substance - identity as constituted \textit{through} relation rather than prior to it \cite{emirbayer1997manifesto}. Other traditions resist the individual as a meaningful unit of analysis altogether. Each implies different answers to questions simulation designers rarely ask explicitly.

The unigraph offers one instantiation of what it looks like to build from a different set of answers - specifically, one that conceives the group as a unified structure through which individuals are constituted rather than a mere aggregate of atomic individuals. As Winograd and Flores argue, our design shapes what can be thought and done within a system \cite{winograd1986understanding}, and this is especially consequential in generative agent simulation where the object of analysis is the social world in all its messiness. The assumptions we encode about personhood, memory, collective life and beyond do not merely shape what our systems can represent but also what we think is worth representing. We invite the HCI community to ask what interaction possibilities might be opened up by alternative ontological commitments. The ontology of generative agent simulation is an open design question.

\section{Conclusion}
We presented \textsc{Synonymix}, a pipeline for constructing privacy-preserving synthetic agent banks from life story personas via graph-based abstraction, demonstrating that it preserves behavioral signal beyond demographic-only baselines while producing personas that are composite of many but copies of none. Beyond synthetic data generation, we propose the unigraph as an artifact for \textit{meso-level simulation} - enabling interaction with group-level representations grounded in individual-level nuance. This opens design questions we invite the CHI community to explore: What modalities might meso-level simulations enable? Can life story personas adequately capture the texture of lived realities? What alternative ontological \textit{frames} can simulations adopt, and what does each highlight and obscure? And, where \textit{should} generative agent simulation head if we move beyond predictive accuracy - beyond treating individuals as "high-fidelity" data points that are data points nonetheless?

\begin{acks}
We are grateful to Joon Sung Park, Helena Vasconcelos, Carolyn Zou, and Jonne Kamphorst for their advice and feedback during various phases of this work. We thank the reviewers for their feedback and borrow their vocabulary to delineate the capabilities and limitations of \textsc{Synonymix}. LLMs were utilized for (1) coding the graphs and tables, and (2) editing and typesetting the written product (text, citations, etc.) of this work.
\end{acks}

\bibliographystyle{ACM-Reference-Format}
\bibliography{references}
\appendix

\section{Edge Grammar}
\label{edge_grammar}
Edges formalize relations between nodes, restricted to four types: \textit{F→S}, \textit{F→F}, \textit{F→I}, and \textit{I→F}. 
\begin{itemize}
    \item \textbf{F->S: Spatial, temporal, and relational markers.} Example edge labels: \textit{by, of, at, in, for, to, from, during, using, with, via, on}). These labels are standardized via a \textit{role registry} (see below).
    \item \textbf{F->F: Temporal or causal relations between concrete events.} Allowed edge labels: "\textit{precedes}" (chronology only), "\textit{enables}" (prerequisite/resource relationship), "\textit{causes}" (explicit causal implication)
    \item \textbf{F->I: }\textbf{Derived meanings from factual events.} Allowed edge labels: "\textit{yields}" (factual event produces or gives rise to an enduring interpretation (belief/value/stance)), "\textit{evokes}" (factual event triggers a transient interpretive reaction), "\textit{supports}" (factual event provides evidence that reinforces an existing interpretation)
    \item \textbf{I->F: Values/Beliefs influencing actions.} Allowed edge labels: "\textit{guides}" (interpretation provides direction or motivation that leads to the factual outcome), "\textit{constrains}" (interpretation imposes limits/conditions that shape the outcome)
\end{itemize}

\subsection{F→S Role Registry}
For F→S edges, we role-type the participants and contextual anchors of factual events. Our role inventory builds on the argument-structure conventions of PropBank \cite{palmer2005proposition} and the contextual frame semantics of FrameNet \cite{baker1998berkeley}, but adapts them to the autobiographical domain by introducing interpretive nodes and restricting the set of allowable edge types. The allowed roles include: \textit{AGENT} (the initiator of an action), \textit{PATIENT} (the entity acted upon), \textit{LOCATION} (the place where the event occurred), \textit{ORGANIZATION} (the institution involved), \textit{DISCIPLINE} (the activity domain or field of practice), \textit{INSTRUMENT} (a tool or enabling method), \textit{RECIPIENT} (the beneficiary of the action), and \textit{TIME} (the time period during which the event occurred). While \textit{TIME} is not typically represented as a subject in linguistic role inventories, we include it here both to anchor autobiographical events chronologically and to support resampling during graph traversal. Attaching temporal markers as roles rather than embedding them in factual node labels allows them to be flexibly perturbed or reassigned during unigraph sampling, thereby obscuring sensitive details (e.g., shifting the period of a life event) while preserving aggregate temporal patterns.

\section{GSS 2022 Core - Example Items}
\label{GSS_2022}
\subsection{Ordinal Items}
\begin{lstlisting}[basicstyle=\ttfamily\small, breaklines=true, columns=fullflexible]

{
    NATSPACY: {
        "question": "Are we spending too much, too little, or about the right amount on Space exploration?",
        "options": {
            "1": "TOO LITTLE",
            "2": "ABOUT RIGHT",
            "3": "TOO MUCH"
        },
        "DEMOGRAPHIC": false,
        "ordinal": true,
        "options_count": 3,
    },
    "DISCAFFWV}": {
        "question": "What do you think the chances are these days that a woman won't get a job or 
        promotion while an equally or less qualified man gets one instead. Is this very likely, 
        somewhat likely, somewhat unlikely, or very unlikely these days?",
        "options": {
            "1": "VERY LIKELY",
            "2": "SOMEWHAT LIKELY",
            "3": "NOT VERY LIKELY",
            "4": "VERY UNLIKELY"
        },
        "DEMOGRAPHIC": false,
        "ordinal": true,
        "options_count": 4
    },
    "CONFED": {
        "question": "As far as the people running the EXECUTIVE BRANCH OF THE FEDERAL GOVERNMENT in 
        this country are concerned, would you say you have a great deal of confidence, only some 
        confidence, or hardly any confidence at all in them?",
        "options": {
            "1": "A GREAT DEAL",
            "2": "ONLY SOME",
            "3": "HARDLY ANY"
        },
        "DEMOGRAPHIC": false,
        "ordinal": true,
        "options_count": 3
    },
    "HAPMAR": {
        "question": "Taking things all together, how would you describe your marriage? Would you say 
        that your marriage is very happy, pretty happy, or not too happy?",
        "options": {
            "1": "VERY HAPPY",
            "2": "PRETTY HAPPY",
            "3": "NOT TOO HAPPY"
        },
        "DEMOGRAPHIC": false,
        "ordinal": true,
        "options_count": 3
    },
    "WLTHHSPS": {
        "question": "On a scale of 1 to 7 with 1 being 'rich' and 7 being 'poor', with 4 being 
        'somewhere in between', do you think people in this group tend to be rich or poor?: Hispanic 
        Americans?",
        "options": {
            "1": "1 - RICH",
            "2": "2",
            "3": "3",
            "4": "4",
            "5": "5",
            "6": "6",
            "7": "7 - POOR"
        },
        "DEMOGRAPHIC": false,
        "ordinal": true,
        "options_count": 7
    },
\end{lstlisting}

\subsection{Nominal Items}
\begin{lstlisting}[basicstyle=\ttfamily\small, breaklines=true, columns=fullflexible]

{
    "CAPPUN": {
        "question": "Do you favor or oppose the death penalty for persons convicted of murder?",
        "options": {
            "1": "FAVOR",
            "2": "OPPOSE"
        },
        "DEMOGRAPHIC": false,
        "ordinal": false,
        "options_count": 2
    },
    "SUICIDE2": {
        "question": "Do you think a person has the right to end his or her own life if: This person 
        has gone bankrupt?",
        "options": {
            "1": "YES",
            "2": "NO"
        },
        "DEMOGRAPHIC": false,
        "ordinal": false,
        "options_count": 2
    },
    "ABPOOR": {
        "question": "Please tell me whether or not you think it should be possible for a pregnant 
        woman to obtain a legal abortion if: The family has a very low income and cannot afford any 
        more children?",
        "options": {
            "1": "YES",
            "2": "NO"
        },
        "DEMOGRAPHIC": false,
        "ordinal": false,
        "options_count": 2
    }
\end{lstlisting}

\section{Demographic and Life Story Personas}
\label{personas}
\subsection{Demographic Persona Example }
"I am a 26-year-old male. I identify as black or african american and am not Hispanic. I was born in the United States. I live in a medium size central city in Middle Atlantic America (New York, New Jersey, Pennsylvania).",
    "I am never married. I have no children. I completed 4 years of college and attained a bachelors degree. I studied optometry.",
    "I work full-time. I am employed by a private employer, not the government. I have a supervisor and also supervise others. My family income last year was $90,000 to $109,999, which I consider average. My personal earnings were $50,000 to $59,999. I identify as working class. In the last ten years, I have not been unemployed for more than a month. When I worked in recent years, it was typically full-time. I have no military service.",
    "I have no religious preference and was raised protestant. Politically, I identify as independent. When I was 16, I lived with my mother only in a small city or town (under 50,000) in the middle atlantic region. At that time, I was living in same state, same city. My family income at that time was below average. My mother completed 12th grade. She worked for pay while I was growing up. Regarding my parents' birthplaces: mother yes, father no. I have four siblings."

\subsection{Life Story Persona (With McAdam's Prompt Guidance)}
    "I grew up in a cramped two‑bedroom apartment on the north side of a mid‑Atlantic city, the kind of place where the hallway always smelled faintly of laundry detergent and the distant rumble of the commuter train could be heard through the thin walls. I was eight, perched on the cracked kitchen floor, watching my mother iron a shirt while the radio played an old Motown record. The heat from the iron warmed my back, and the scent of fresh laundry mixed with the faint grease of the fried chicken we’d just eaten. When she glanced up and said, “You’re the only one who can keep the lights on for us,” I felt a sudden weight settle in my chest—a mix of pride and the pressure of being the one who had to help. That night I promised myself I would finish school, no matter how hard it got.",
    "At fifteen, I was standing in the cramped hallway of the public high school’s science lab, the fluorescent lights buzzing overhead, the smell of chemicals sharp in the air. My biology teacher, Mr. Alvarez, handed me a slide of a rabbit’s eye and asked, “What do you see?” I felt my heart race, the buzz of the fluorescent tubes matching the buzz in my ears. I whispered, “A future,” and he smiled, “Then you’ll need to see farther.” That moment nudged me toward a science track, even though my mother, who had only a high‑school diploma, warned that “college is a luxury.” I took that warning as a challenge, not a barrier.",
    "When I was twenty‑two, I walked into the bustling downtown optometry clinic where I had just been hired as an assistant. The waiting room smelled of antiseptic and cheap coffee; the hum of the auto‑refractor filled the space. My supervisor, a woman with a scar across her cheek, handed me a stack of patient charts and said, “You’re the first Black guy they’ve hired in years—don’t let them forget why you’re here.” My palms sweated, but the click of the keyboard under my fingers felt like a promise. I spent the next year studying for the optometry licensing exam, pulling all‑nighters in a tiny studio apartment where the radiator hissed like a low‑wail. Passing that exam taught me that the path to a stable paycheck could be forged with relentless, quiet work.",
    "The first real crisis hit when I was twenty‑six, three months into my new role as a licensed optometrist at a private practice. I was in the break room, the fluorescent light flickering, the smell of burnt coffee thick, when the office manager called me into her glass‑walled office. She stared at me over her glasses, voice flat, “We have to cut staff; your position is being eliminated.” The words hit like a cold draft. I felt my stomach drop, the taste of metal in my mouth, and the fear that the rent I shared with my partner, Maya, would slip through our fingers. I left the room clutching a stack of patient files, the weight of the future pressing against my shoulders. That day I learned that security could evaporate in an instant, and I resolved to build a safety net, even if it meant picking up extra shifts at the local urgent‑care center.",
    "Two weeks later, I was at home in our modest two‑bedroom apartment, the kitchen light buzzing, the scent of Maya’s jasmine shampoo filling the air. We sat at the kitchen table, a stack of unpaid bills spread before us, the rent notice glaring in red ink. I could feel the paper’s rough texture against my fingertips, hear the ticking of the wall clock, and sense Maya’s hand trembling in mine. We talked about taking a second job, about cutting back on groceries, about the fear that my student loans would never disappear. The moment cemented a new resolve: I would start a side business offering vision screenings at community centers, using the equipment I’d saved from the clinic. It taught me that survival sometimes means turning a setback into a side hustle.",
    "Now, at twenty‑six, I’m working full‑time at a private optometry practice, supervising two technicians, while living with Maya in the same house we’ve made our own. My income sits at about \$57,000 a year, which, after rent, utilities, and Maya’s student loans, leaves little room for anything beyond the essentials. I spend my evenings after the last patient leaves, the clinic lights dimming, the scent of disinfectant lingering, scrolling through insurance forms and budgeting spreadsheets. I’m constantly weighing whether to apply for a loan to buy a second set of lenses for my side business or to put a little extra toward a down‑payment on a house. The future feels like a tightrope, but the rope is my own making, and I’m still learning how to keep my balance."

\section{Persona Method Discussion}
\label{persona_method}
For persona generation and evaluation pipeline, we call \textit{gpt-oss-120b} model hosted on Lightning.AI for inference due to compute constraints.

\subsection{Generating Life Story Persona (L)}
An issue we noticed during synthetic life story generation is that while the facts of the life stories are different, the structure and linguistic style is remarkably similar. For example, multiple personas used the same linguistic markers of the "fluorescent lamp" and the "kitchen floor", which we attribute to LLM stylistic mode collapse.

To mitigate this, we augmented the persona generation method with verbalized sampling - a method for sampling response outputs from the tail end of the probability distribution \cite{zhang2025verbalizedsamplingmitigatemode}. However, similarly low-probability life story outputs still consistently retained similar stylistic markers. While this does not affect our evaluation's test for internal validity, we would note this here as a core limitation of LLM generated synthetic dataset.

\end{document}